\documentclass[12pt]{iopart}

 \usepackage{graphicx}

\begin{document}

\title[Parabolic odd potential]{One-dimensional quantum scattering \\ by a parabolic odd potential}
\author{E M Ferreira$^1$ and J Sesma$^2$}
\address{$^1$ Instituto de F\'{\i}sica, Universidade Federal do Rio
de Janeiro, 21941-972, Rio de Janeiro, Brasil}
\address{$^2$ Departamento de F\'{\i}sica Te\'orica, Facultad de
Ciencias, 50009, Zaragoza, Spain}

\eads{\mailto{erasmo@if.ufrj.br}, \mailto{javier@unizar.es}}

\begin{abstract}
Quantum scattering by a one-dimensional odd potential proportional to the square of the distance to the origin is considered. The Schr\"odinger equation is solved exactly and explicit algebraic expressions of the wave function are given. A complete discussion of the scattering function reveals the existence of Gamow (decaying) states and of resonances.
\end{abstract}

\pacno{03.65.Nk; 02.30.Hq}




\section{Introduction}

An amazing property of divergent-at-infinity odd one-dimensional potentials of the type
\begin{equation}
V(x) = -\,x^N, \qquad N=3, 5, 7, \ldots, \label{i1}
\end{equation}
or
\begin{equation}
V(x)=\left\{\begin{array}{lll}x^N,&\qquad \mbox{for}&\quad x<0, \\ -\,x^N, &\qquad \mbox{for}&\quad x>0.\end{array}\right.\qquad N= 4, 6, 8, \ldots, \label{i2}
\end{equation}
is their capability to sustain resonances. The first evidence of that property did not occur in a direct way, but in the study of a Hamiltonian where a term of the type of Eq. (\ref{i1}), with $N=3$, had been added to the familiar harmonic oscillator, to have what is known as cubic anharmonic oscillator. Old studies \cite{yari,alva1} of that Hamiltonian,
\begin{equation}
H=-\,\frac{d^2}{dx^2}+\frac{x^2}{4}-\lambda\,x^3, \qquad \lambda>0,  \label{i3}
\end{equation}
revealed that it has complex eigenvalues corresponding to localized eigenfunctions, that is, eigenstates of complex energy that could be associated with resonances. Investigations of different aspects of these resonances have continued up to recent years \cite{alva2,jent1}.  The existence of localized eigenstates was initially attributed to the potential barrier due to the presence of the term $x^2$. But progressive weakening of that term does not destroy the resonances, which are present even in a pure cubic potential $V(x)=-x^3$ \cite{jent1}. This fact makes it interesting to study resonances in potentials of the form given in Eqs. (\ref{i1}) and (\ref{i2}), without the presence of a harmonic oscillator term.

On the other hand, resonances are not possible in a linear potential, i. e., a potential as given by Eq. (\ref{i1}) with $N=1$, a fact already mentioned in \cite{alva1}. There are, in this case, no privileged values of the energy. This fact becomes obvious if one considers that a displacement of the energy accompanied by a corresponding translation in the variable $x$ leaves invariant the eigenvalue problem.

The question arises if a potential with a shape, shown in Fig. 1, intermediate between those of the linear and cubic potentials, namely the parabolic odd one,
\begin{equation}
V(x)=\left\{\begin{array}{lll}x^2,&\qquad \mbox{for}&\quad x<0, \\ -\,x^2, &\qquad \mbox{for}&\quad x>0,\end{array}\right.  \label{i4}
\end{equation}
which is of the form of Eq. (\ref{i2}) with $N=2$, can sustain resonances. Besides, one-dimensional scattering by this parabolic odd potential presents the additional interest of being algebraically solvable. These circumstances have lead us to carry out, in the present paper, a thorough study of the scattering by the potential given in Eq. (\ref{i4}).
\begin{figure}
\begin{center}
\vspace{1cm}\includegraphics{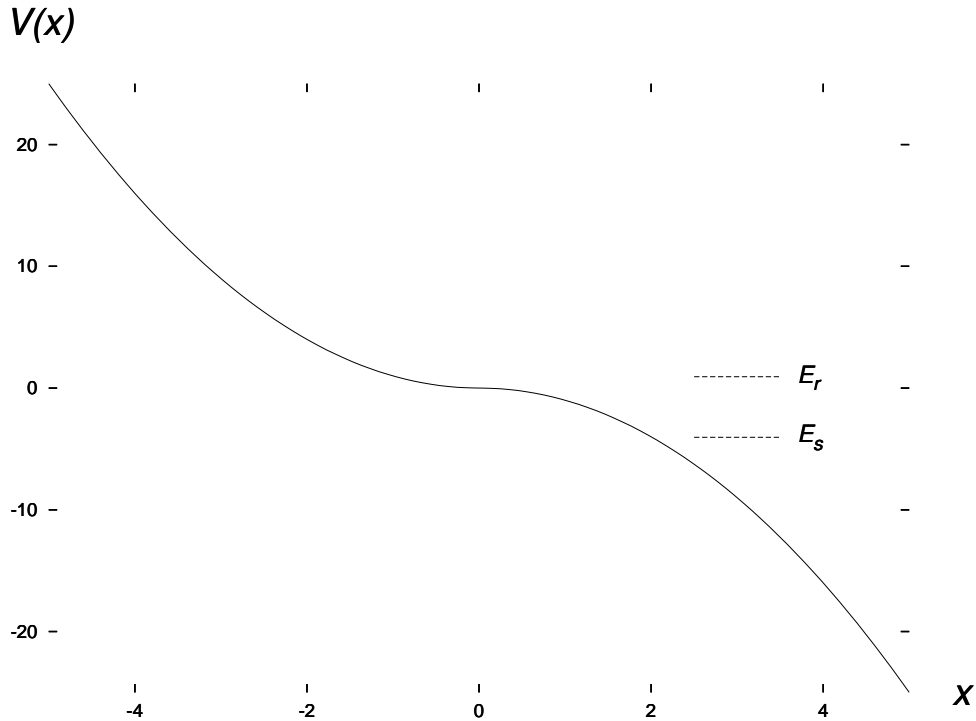}
\end{center}
\caption{Graphical representation of the potential given by Eq. (\ref{i4}). The horizontal dashed lines correspond to two particular values of the energy, $E_r=0.935$ and $E_s=-4.042626$ to be considered in Section 6.}
\label{fig:1}
\end{figure}

The problem to be discussed here possesses common features with the study, done by Barton \cite{bart}, of tunneling and scattering in the inverted oscillator (also known as parabolic barrier). We adopt here the pragmatic attitude of Barton, leaving aside the mathematical issues mentioned in \cite{bart} and addressed in more recent papers \cite{bala}.

Scattering by one-dimensional finite-range potentials constitutes a chapter in most texts of Quantum Mechanics. (See, for instance, the classical book by Landau and Lifshitz \cite[Section 22]{land} or the more recent ones by Robinett \cite[Chapter 12]{robi} and  by Newton \cite[Chapter 3]{newt}. For a recent review, see \cite{boya}.) Assuming a probability flux impinging from the right on a potential that vanishes out of the interval $[-a,a]$, the wave function in the outer region can be written in the form
\begin{equation}
\psi(x) = \left\{\begin{array}{ll}\psi^{\rm{incoming}}(x) + r\,\psi^{\rm{outgoing}}(x), & \quad x>a, \\ t\,\psi^{\rm{outgoing}}(x), & \quad x<-\,a. \end{array}\right.  \label{i5}
\end{equation}
The  reflection and transmission coefficients, respectively  $r$ and $t$,
are functions of the energy of the particle represented by the incident flux.
In the case of the potential becoming infinite and, therefore, impenetrable, the last
equation is replaced by
\begin{equation}
\psi(x) = \left\{\begin{array}{ll}\psi^{\rm{incoming}}(x) - S\,\psi^{\rm{outgoing}}(x), & \quad x>a, \\ 0, &\quad x<-\,a, \end{array}\right.  \label{i6}
\end{equation}
where the  the scattering coefficient  $S$  is also dependent on the energy.
The potential of Eq. (\ref{i4}) that we are going to consider here  has an
infinite range. Nevertheless, the preceding formalism is applicable. Equations (\ref{i5}) and (\ref{i6}) remain valid if one replaces $x>a$ and $x<-a$ respectively by $x\to +\infty$ and $x\to -\infty$ and uses adequate expressions \cite{bart} for $\psi^{\rm{incoming}}(x)$ and $\psi^{\rm{outgoing}}(x)$.

In Section 2 we obtain the Frobenius and Thom\'e solutions of the Schr\"odinger equation with
the potential defined in Eq. (\ref{i4}). The connection factors linking these two kinds of solutions are obtained in Section 3. In this way we are able to write the physical solution and the scattering function in Section 4. A study, in Section 5, of the analytic properties of the scattering function in the complex energy plane reveals the occurrence of Gamow states, whose correspondence with resonances is discussed in Section 6. Finally, Section 7 contains some comments about the peculiarities of the potential considered.

\section{Solutions of the Schr\"odinger equation for the parabolic odd potential}

The differential equation to be solved is (in adequate scales for lengths and energies)
\begin{equation}
-\,\frac{d^2\psi(x)}{dx^2}+V(x)\,\psi(x)=E\,\psi(x),  \label{i7}
\end{equation}
with $V(x)$ given by Eq. (\ref{i4}). Let us start by solving it on the positive real semiaxis. Written for $x>0$, Eq. (\ref{i7}) becomes
\begin{equation}
\frac{d^2\psi(x)}{dx^2}+(x^2+E)\psi(x)=0,  \qquad x>0.\label{i8}
\end{equation}
Two convergent power series solutions (Frobenius ones) can be immediately obtained. However, for convenience in connecting the solutions valid at small $x$ to their asymptotic form, we express the Frobenius solutions as a product of an exponential times a convergent series, that turns out to be a confluent hypergeometric function. One gets in this way
\begin{eqnarray}
\psi_1^+(x)&=&\exp(ix^2/2)\,\ _1F_1\left(\frac{1-iE}{4}; \frac{1}{2}; -ix^2\right),  \label{i9} \\
\psi_2^+(x)&=&x\,\exp(ix^2/2)\,\ _1F_1\left(\frac{3-iE}{4}; \frac{3}{2};-i x^2\right). \label{i10}
\end{eqnarray}
Formal solutions expressed as the product of an exponential times an asymptotic expansion (Thom\'e solutions) for $x\to +\infty$ can also be obtained by substitution in the differential equation. They are, in terms of generalized hypergeometric functions,
\begin{eqnarray}
\psi_3^+(x)&=&\exp(ix^2/2)\,x^{-(1-iE)/2}\,\ _2F_0\left(\frac{1-iE}{4}, \frac{3-iE}{4};; -\frac{i}{x^2}\right),  \label{i11} \\
\psi_4^+(x)&=&\exp(-ix^2/2)\,x^{-(1+iE)/2}\,\ _2F_0\left(\frac{1+iE}{4}, \frac{3+iE}{4};; \frac{i}{x^2}\right).  \label{i12}
\end{eqnarray}

Let us now consider the solutions on the negative real semiaxis, $x<0$. The differential equation is now
\begin{equation}
\frac{d^2\psi(x)}{d^2x}+(-x^2+E)\psi(x)=0,  \qquad x<0.\label{i13}
\end{equation}
Following the same procedure as for Eqs. (\ref{i9}) to (\ref{i12}), we  find for the Frobenius solutions
\begin{eqnarray}
\psi_1^-(x)&=&\exp(-x^2/2)\,\ _1F_1\left(\frac{1-E}{4}; \frac{1}{2}; x^2\right),  \label{i14} \\
\psi_2^-(x)&=&x\,\exp(-x^2/2)\,\ _1F_1\left(\frac{3-E}{4}; \frac{3}{2}; x^2\right),   \label{i15}
\end{eqnarray}
and for the Thom\'e solutions
\begin{eqnarray}
\psi_3^-(x)&=&\exp(-x^2/2)\,x^{-(1-E)/2}\,\ _2F_0\left(\frac{1-E}{4}, \frac{3-E}{4};; -\frac{1}{x^2}\right),  \label{i16} \\
\psi_4^-(x)&=&\exp(x^2/2)\,x^{-(1+E)/2}\,\ _2F_0\left(\frac{1+E}{4}, \frac{3+E}{4};; \frac{1}{x^2}\right).  \label{i17}
\end{eqnarray}

It is immediate to check that $\psi_j^-$ and $\psi_j^+$  ($j=1,2$) take the same value at $x=0$. The same is true for their derivatives with respect to $x$. Therefore, we have obtained two solutions of Eq. (\ref{i7}), $\psi_j(x)$ ($j=1,2$), which are represented by $\psi_j^-(x)$ when $x\leq 0$ and by $\psi_j^+(x)$ if $x\geq 0$. Since these two Frobenius solutions constitute a fundamental set of solutions, any other one can be written as a linear combination of them. In particular, the physical solution would be
\begin{equation}
\psi_{\rm{phys}}(x) = A_1\,\psi_1(x)+A_2\,\psi_2(x),  \label{i18}
\end{equation}
with coefficients $A_1$ and $A_2$ to be determined.

\section{The connection factors}

The behavior  of the Frobenius solutions for $x\to +\infty$ can be written in terms of the Thom\'e ones, by means of the so called connection factors $T_{j,k}^+$, in the form
\begin{equation}
\psi_j^+(x) \sim T_{j,3}^+\,\psi_3^+(x) + T_{j,4}^+\,\psi_4^+(x), \qquad j=1,2, \qquad x\to +\infty. \label{i19}
\end{equation}
These connection factors are obtained immediately by using the asymptotic power series of the confluent hypergeometric function \cite[Sec. 2.5, Eq. (47)]{ding}
\begin{eqnarray}
\fl \ _1F_1(a;c;z) &\sim \frac{\Gamma(c)\,z^{a-c}\,e^z}{\Gamma(a)}\ _2F_0(1-a,c-a;;z^{-1})  \nonumber \\
\fl & + \left(\!\begin{array}{c}e^{i\pi a} \\ e^{-i\pi a}\\ \cos \pi a \end{array}\!\right)\frac{\Gamma(c)\,z^{-a}}{\Gamma(c-a)}\ _2F_0(a,a\!-\!c\!+\!1;;(-z)^{-1}), \quad \left.\begin{array}{l}0<\arg z<\pi \\ 0>\arg z>-\pi \\
\arg z=0\end{array} \right\}. \label{i20}
\end{eqnarray}
They turn out to be
\begin{eqnarray}
T_{1,3}^+ =  \frac{e^{-i\pi(1-iE)/8}\,\Gamma(1/2)}{\Gamma((1+iE)/4)}, \qquad && T_{1,4}^+ =  \frac{e^{i\pi(1+iE)/8}\,\Gamma(1/2)}{\Gamma((1-iE)/4)}, \label{i21}  \\
T_{2,3}^+ =  \frac{e^{-i\pi(3-iE)/8}\,\Gamma(3/2)}{\Gamma((3+iE)/4)}, \qquad && T_{2,4}^+ =  \frac{e^{i\pi(3+iE)/8}\,\Gamma(3/2)}{\Gamma((3-iE)/4)}. \label{i22}
\end{eqnarray}

Analogously to Eq. (\ref{i19}), one can express the behavior of the Frobenius solutions for $x\to -\infty$ in terms of the Thom\'e ones,
\begin{equation}
\psi_j^-(x) \sim T_{j,3}^-\,\psi_3^-(x) + T_{j,4}^-\,\psi_4^-(x), \qquad j=1,2, \qquad x\to -\infty. \label{i23}
\end{equation}
For an easier determination of their connection factors, we rewrite the solutions on the negative real semiaxis, that is, for $x=e^{i\pi}|x|$, in the form
\begin{eqnarray}
\psi_1^-(x)&=\exp(-|x|^2/2)\,\ _1F_1\left(\frac{1-E}{4}; \frac{1}{2}; |x|^2\right),  \nonumber \\
\psi_2^-(x)&= - \exp(-|x|^2/2)\,\,|x|\,\ _1F_1\left(\frac{3-E}{4}; \frac{3}{2}; |x|^2\right),  \nonumber \\
\psi_3^-(x)&=\exp(-|x|^2/2)\,e^{-i\pi(1-E)/2}\,|x|^{-(1-E)/2}\,\ _2F_0\left(\frac{1-E}{4}, \frac{3-E}{4};; -\frac{1}{|x|^2}\right),  \nonumber \\
\psi_4^-(x)&=\exp(|x|^2/2)\,e^{-i\pi(1+E)/2}\,|x|^{-(1+E)/2}\,\ _2F_0\left(\frac{1+E}{4}, \frac{3+E}{4};; \frac{1}{x^2}\right).  \nonumber
\end{eqnarray}
Then, by using Eq. (\ref{i20}), we obtain for the connection factors on the negative real semi-axis the expressions
\begin{eqnarray}
\fl T_{1,3}^- =  \frac{e^{i\pi(1-E)/2}\,\cos((1-E)\pi/4)\,\Gamma(1/2)}{\Gamma((1+E)/4)}, \quad && T_{1,4}^- =  \frac{e^{i\pi(1+E)/2}\,\Gamma(1/2)}{\Gamma((1-E)/4)}, \label{i24}  \\
\fl T_{2,3}^- = -\,\frac{e^{i\pi(1-E)/2}\,\cos((3-E)\pi/4) \,\Gamma(3/2)}{\Gamma((3+E)/4)}, \quad && T_{2,4}^- = -\, \frac{e^{i\pi(1+E)/2}\,\Gamma(3/2)}{\Gamma((3-E)/4)}. \label{i25}
\end{eqnarray}

\section{The scattering function}

We have, already, all we need to calculate the coefficients $A_1$ and $A_2$ in the expression of the physical solution, Eq. (\ref{i18}). In view of Eq. (\ref{i23}), one has for $x\to -\infty$
\begin{equation}
\psi_{\rm{phys}}(x) \sim (A_1\,T_{1,3}^-+A_2\,T_{2,3}^-)\,\psi_3^-(x)+(A_1\,T_{1,4}^-+A_2\,T_{2,4}^-)\,\psi_4^-(x).  \label{i26}
\end{equation}
The potential barrier $x^2$ prevents the hypothetical particle represented by $\psi_{\rm{phys}}(x)$ to reach large negative values of $x$. Therefore, the diverging (for $x\to -\infty)$ component $\psi_4^-$ in the expression of $\psi_{\rm{phys}}(x)$ must be eliminated, that is, the coefficients $A_1$ and $A_2$ must be taken such that
\begin{equation}
A_1\,T_{1,4}^-+A_2\,T_{2,4}^-=0.  \label{i27}
\end{equation}
This relation determines them up to a common arbitrary multiplicative constant, that may be fixed by requiring the fulfilment of an additional condition like, for instance,
\begin{equation}
A_1\,T_{1,4}^++A_2\,T_{2,4}^+=1,  \label{i28}
\end{equation}
unless it happens that
\begin{equation}
T_{1,4}^+T_{2,4}^--T_{2,4}^+T_{1,4}^- = 0,  \label{i29}
\end{equation}
in which case
\begin{equation}
A_1\,T_{1,4}^++A_2\,T_{2,4}^+=0.  \label{i30}
\end{equation}
Leaving aside this case, that will be considered in Section 5, one obtains from Eqs. (\ref{i27}) and (\ref{i28})
\begin{equation}
A_1=\frac{T_{2,4}^-}{T_{1,4}^+T_{2,4}^--T_{2,4}^+T_{1,4}^-},\qquad A_2=\frac{-\,T_{1,4}^-}{T_{1,4}^+T_{2,4}^--T_{2,4}^+T_{1,4}^-}.  \label{i31}
\end{equation}
On the other hand, bearing in mind Eqs. (\ref{i18}), (\ref{i19}), and (\ref{i28}), one realizes that, for $x\to +\infty$,
\begin{equation}
\psi_{\rm{phys}}(x) \sim (A_1\,T_{1,3}^++A_2\,T_{2,3}^+)\,\psi_3^+(x)+\psi_4^+(x).  \label{i32}
\end{equation}
As it is well known, the flux of probability associated to a wave function $\psi(x)$ is given (in appropriate units) by
\begin{equation}
j(x) = -\,i\left(\psi(x)^*\,\frac{d\psi(x)}{dx}-\frac{d\psi(x)^*}{dx}\,\psi(x)\right),  \nonumber
\end{equation}
the asterisk standing for complex conjugation.
It is immediate to check that the fluxes associated with $\psi_3^+(x)$ and $\psi_4^+(x)$, as given by Eqs. (\ref{i11}) and (\ref{i12}), are respectively positive and negative. Besides, for real $E$, $\psi_3$ and $\psi_4$ are complex conjugate to each other and, obviously, their moduli are equal. Consequently, they represent, respectively, an outgoing (to the right) wave and an incoming (from the right) one.
Therefore, Eq. (\ref{i32}) is of the form
\begin{equation}
\psi_{\rm{phys}}(x) \sim -\,S(E)\,\psi^{\rm{outgoing}}(x)+\psi^{\rm{incoming}}(x),  \label{i33}
\end{equation}
analogous to Eq. (\ref{i6}), with a scattering function
\begin{equation}
S(E) = -(A_1\,T_{1,3}^++A_2\,T_{2,3}^+)
     = -\, \frac{T_{1,3}^+T_{2,4}^--T_{2,3}^+T_{1,4}^-}{T_{1,4}^+T_{2,4}^--T_{2,4}^+T_{1,4}^-}.  \label{i34}
\end{equation}
Substitution of the connection factors by their expressions, given in Eqs. (\ref{i21}), (\ref{i22}), (\ref{i24}), and (\ref{i25}), allows one to obtain
\begin{equation}
S(E) =  -\,e^{-i\pi/2}\,\frac{N(E)}{D(E)},   \label{i35}
\end{equation}
where we have denoted
\begin{eqnarray}
N(E) & = & \frac{e^{i\pi/8}}{\Gamma\left(\frac{3-E}{4}\right)\,\Gamma\left(\frac{1+iE}{4}\right)}
+ \frac{e^{-i\pi/8}}{\Gamma\left(\frac{1-E}{4}\right)\,\Gamma\left(\frac{3+iE}{4}\right)},  \label{i36} \\
D(E) & = & \frac{e^{-i\pi/8}}{\Gamma\left(\frac{3-E}{4}\right)\,\Gamma\left(\frac{1-iE}{4}\right)}
+ \frac{e^{i\pi/8}}{\Gamma\left(\frac{1-E}{4}\right)\,\Gamma\left(\frac{3-iE}{4}\right)}.  \label{i37}
\end{eqnarray}

It is evident, from their explicit expressions, that $N(E)$ and $D(E)$ are complex conjugate to each other, as long as $E$ is real. Consequently,
\begin{equation}
|S(E)|=1  \qquad \mbox{for real}\; E.   \label{i39}
\end{equation}
It is therefore possible to describe the result of the scattering in terms of a phase shift $\delta(E)$ defined as usually \cite{kahn}
\begin{equation}
S(E)=\exp \left[2\,i\,\delta(E)\right].  \label{i40}
\end{equation}
This definition of the phase shift is not unambiguous: it determines $\delta(E)$ up to addition of $n\pi$ ($n$ integer). To eliminate that ambiguity, we have chosen the interval $[0, \pi)$ to contain the value of $\delta(0)$. The resulting values of $\delta(E)$, for $-10<E<15$, are represented in Fig. 2.
\begin{figure}
\begin{center}
\vspace{0.5cm}\includegraphics{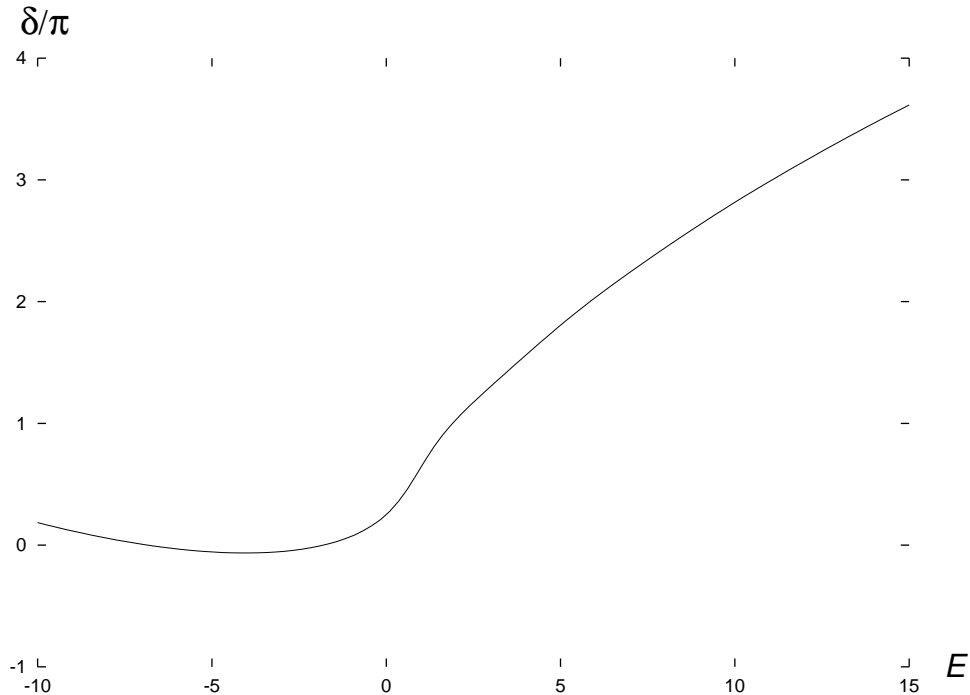}
\end{center}
\caption{Phase shift, in units of $\pi$, of a wave scattered by the parabolic odd potential vs. the energy of the particle represented by the wave.}
\label{fig:2}
\end{figure}

\section{Analytic properties of the scattering function}

In the preceding sections, real values for the energy were implicitly assumed. It is widely recognized that valuable information about the scattering process can be obtained by a study of the analytic properties of the scattering function extended to complex values of the energy. In the present case, such extension does not present any difficulty. The scattering function appears in Eq. (\ref{i35}) as the quotient of two functions, $N(E)$ and $D(E)$, defined in terms of the reciprocal Gamma function, $1/\Gamma(z)$, which can be trivially extended to complex values of $z$. In fact, it admits a series expansion convergent in the whole finite complex $z$-plane,
\begin{equation}
\frac{1}{\Gamma (z)}=\sum_{k=1}^\infty a_k\,z^k,  \label{r1}
\end{equation}
whose coefficients $a_k$, approximated to 31 digits, can be found in a paper by Wrench \cite{wren}. Therefore, given that $N(E)$ is finite for any complex value of $E$, the only singularities of $S(E)$ in the finite complex $E$-plane are due to zeros of its denominator, that is, to fulfilment of Eqs. (\ref{i29}) and (\ref{i30}),
in which case we have, instead of Eq. (\ref{i32}),
\begin{equation}
\psi_{\rm{phys}}(x) \sim (A_1\,T_{1,3}^++A_2\,T_{2,3}^+)\,\psi_3^+(x), \qquad \mbox{for}\quad x\to +\infty .  \label{i41}
\end{equation}
It is not difficult to see that the upper half-plane, $\Im E>0$, is free from such singularities. From Eq. (\ref{i7}) and its complex conjugate, one obtains immediately
\begin{equation}
\fl \frac{d}{dx}\left(\psi_{\rm{phys}}(x)\,\frac{d\psi_{\rm{phys}}^*(x)}{dx}-\psi_{\rm{phys}}^*(x)\frac{d\psi_{\rm{phys}}(x)}{dx}\right) = (E-E^*)\,|\psi_{\rm{phys}}(x)|^2,   \label{r2}
\end{equation}
which integrated from $-\infty$ to $x$ gives
\begin{equation}
\fl\psi_{\rm{phys}}(x)\,\frac{d\psi_{\rm{phys}}^*(x)}{dx}-\psi_{\rm{phys}}^*(x)\frac{d\psi_{\rm{phys}}(x)}{dx} = (E-E^*)\int_{-\infty}^x|\psi_{\rm{phys}}(t)|^2\,dt\,.   \label{r3}
\end{equation}
The right hand side of this equation is pure imaginary, its modulus increases with $x$, and its sign is that of $\Im E$. For the left hand side, assuming $x$ positive and sufficiently large, we have, from Eq. (\ref{i41}) ($\mathcal{W}[f,g]$ representing the Wronskian of the functions $f$ and $g$)
\begin{eqnarray}
\mathcal{W}\left[\psi_{\rm{phys}},\,\psi_{\rm{phys}}^*\right](x) &\sim |A_1\,T_{1,3}^++A_2\,T_{2,3}^+|^2\, \mathcal{W}\left[\psi_3^+,\,(\psi_3^+)^*\right](x) \nonumber \\
&\sim |A_1\,T_{1,3}^++A_2\,T_{2,3}^+|^2\,(-2\,i)\,x^{i(E-E^*)/2},   \label{r4}
\end{eqnarray}
which may present the characteristics of the right hand side of Eq. (\ref{r3}) only if $\Im E<0$.

In the general case of complex energy, one realizes that
\begin{equation}
D(E^*)=\left[(N(E)\right]^* \qquad \mbox{and}\qquad N(E^*)=\left[D(E)\right]^*,  \label{r5}
\end{equation}
from which one obtains the familiar unitarity condition
\begin{equation}
S(E)\,[S(E^*)]^* = 1.  \label{i38}
\end{equation}
According to this property, common to finite range potentials, zeros of $S(E)$ appear in the upper half-plane at positions symmetrical with respect to the horizontal axis of those of the poles in the lower half plane. There is, however, a symmetry property of zeros and poles of the scattering function specific of the potential we are considering. It stems from the relations
\begin{equation}
N(iE^*)= [N(E)]^*,\qquad D(-iE^*)=[D(E)]^*, \label{i43}
\end{equation}
that can be trivially checked in  Eqs. (\ref{i36}) and (\ref{i37}). Such relations imply that the pattern of zeros of $S(E)$ is symmetrical with respect to the bisector of the first and third quadrants in the $E$-plane, and the pattern of poles is symmetrical with respect to the bisector of the second and fourth quadrants. This symmetry, together with the impossibility of having poles in the upper half-plane proven above, allows one to conclude that, for the potential we are considering, poles of the scattering function  may occur only in the fourth quadrant of the $E$-plane.

All these analytic properties of the scattering function are confirmed by a numerical computation of $S(E)$ as given by Eqs. (\ref{i35}), (\ref{i36}) and (\ref{i37}). We have represented in Fig. 3 the modulus and phase of the scattering function for complex values of $E$ in the region $-10\leq\Re E\leq15$ and $-15\leq\Im E\leq 0$. In view of the unitarity condition, Eq. (\ref{i38}), it is sufficient to show $S(E)$ in the lower half-plane. Constant-phase lines are symmetrical with respect to the horizontal axis, $\Im E=0$. Constant-modulus lines associated with  $|S|=a$ in the lower half-plane and $|S|=1/a$ in the upper one are also symmetric to each other. In the figure, the constant-modulus (solid) lines are labeled with the value of $\log_{10}[|S(E)|]$. The shown constant-phase (dashed) lines correspond to $\arg [S(E)]=n\,\pi/4$ ($n$ integer). With the convention adopted above to remove the ambiguity in the phase shift, lines  intersecting the horizontal axis correspond, from left to right, to $\arg [S(E)]=\pi/4,\, 0,\, 0,\, \pi/4,\, \pi/2,\, 3\pi/4,\, \pi,\, 5\pi/4,\, \ldots$. Poles of $S(E)$ are immediately recognized in the chart. There seems to exist two infinite sequences of poles, symmetric with respect to the bisector of the fourth quadrant, besides a pole at the bisector. The approximate positions of this pole and the first few of the two infinite sequences are given in Table 1. Although not explicitly shown, one can guess, looking at the figure, the existence of saddle points about positions intermediate between those of two consecutive poles of the same sequence. There is, however, a more interesting saddle point: that on the real axis, at $E\approx -4.042626$, where two $|S|=1$ lines intersect. The two constant-phase lines intersecting also there correspond to $\arg (S) \approx -\,0.519712\,\pi/4$.

\begin{figure}
\begin{center}
\includegraphics{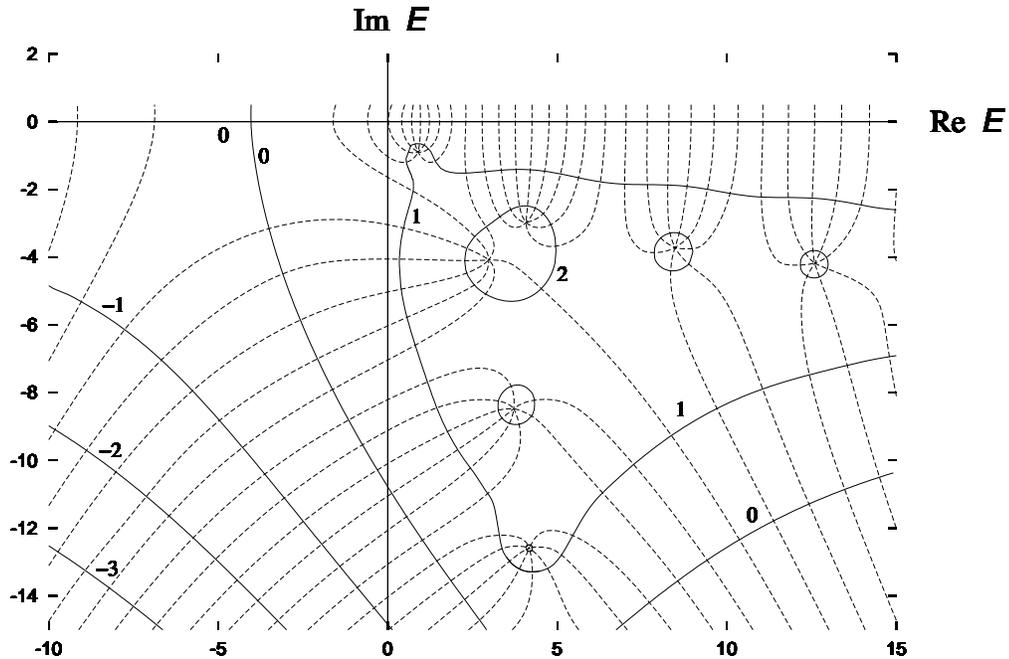}
\end{center}
\caption{Chart of modulus and phase of the scattering function of the parabolic odd potential. See the text for an explanation of the values corresponding to the constant-modulus (solid) lines and constant-phase (dashed) lines.}
\label{fig:3}
\end{figure}

\begin{table}
\caption{Approximate positions, in the complex $E$-plane, of the first poles of the scattering function of the parabolic odd potential.}
\begin{indented}
\item[]
\begin{tabular}{r|r}
\br
\multicolumn{2}{c}
{$0.889605-i\,0.889605$}   \\
$2.977506-i\,4.081280 \quad$ & $ \quad 4.081280-i\,2.977506$ \\
$3.715766-i\,8.472130 \quad$ & $ \quad 8.472130-i\,3.715766$ \\
$4.173994-i\,12.59206 \quad$ & $ \quad 12.59206-i\,4.173994$ \\
$4.509353-i\,16.66338 \quad$ & $ \quad 16.66338-i\,4.509353$ \\
\br
\end{tabular}
\end{indented}
\end{table}

In the next section we will see that the pole of $S(E)$ at the bisector of the fourth quadrant is of special relevance. Approximations to its position can be obtained from polynomial approximations to the equation
\begin{equation}
D(E)=0  ,  \label{r5}
\end{equation}
obtained by truncation of the Taylor expansion
\begin{equation}
D(E)=\sum_{m=0}^\infty b_m\,E^m,  \label{r6}
\end{equation}
where
\begin{equation}
b_m=\frac{1}{m!}\,\left.\frac{d^mD(E)}{dE^m}\right|_{E=0}. \label{r7}
\end{equation}
Trivial calculus gives
\begin{equation}
b_m = \left(-\frac{1}{4}\right)^m\,\sum_{n=0}^m \left( e^{-i\pi/8}\,i^{m-n}+
e^{i\pi/8}\,i^n\right)\frac{G^{(n)}(3/4)\,G^{(m-n)}(1/4)}{n!\,(m-n)!}, \label{r8}
\end{equation}
where we have used the notation $G^{(n)}(z)$ for the successive derivatives of the reciprocal Gamma function,
\begin{equation}
G^{(n)}(z)\equiv\frac{d^n}{dz^n}\frac{1}{\Gamma (z)}, \label{r9}
\end{equation}
whose computation was discussed in a former paper \cite[Appendix B]{abad}. We show, in Table 2, the first solution of
\begin{equation}
\sum_{m=0}^M b_m\,E^m=0  \label{r10}
\end{equation}
for several successive values of $M$.

\begin{table}
\caption{Successive approximations, $E_M$, to the position of the pole of $S(E)$ at the bisector of the fourth quadrant. They have been obtained by solving Eq. (\ref{r10}) with increasing values of $M$. The coefficients $b_m$ of the Taylor expansion of $D(E)$, Eq. (\ref{r6}), are shown in the second column.}
\begin{indented}
\item[]
\begin{tabular}{rll}
\br
 $M$ & $\qquad\qquad b_M $ & $\quad \qquad E_M $  \\
\mr
0 & \quad 0.4158919086E$+$00 & \   \\
1 & \quad 0.3438700716E$+$00\,($-1-i$) & \quad 0.6047224563\,($1-i$) \\
2 & \quad 0.1302850455E$+$00\,\,$i$ & \quad 0.9382649623\,($1-i$) \\
3 & \quad 0.1693998465E$-$02\,($1-i$) & \quad 0.9131374180\,($1-i$) \\
4 & \quad 0.2314470246E$-$02 & \quad 0.8885559742\,($1-i$) \\
5 & \quad 0.4198228545E$-$04\,($-1-i$) & \quad 0.8892565969\,($1-i$) \\
6 & \quad 0.2368862531E$-$04\,($-i$) & \quad 0.8896106601\,($1-i$) \\
7 & \quad 0.5531769758E$-$07\,($-1+i$) & \quad 0.8896091851\,($1-i$) \\
8 & \quad 0.1623934529E$-$06\,($-1$) & \quad 0.8896053333\,($1-i$) \\
9 & \quad 0.3336310538E$-$08\,($-1-i$) & \quad 0.8896051925\,($1-i$) \\
10 & \quad 0.5905347326E$-$09\,\,$i$ & \quad 0.8896052147\,($1-i$) \\
11 & \quad 0.2651914365E$-$10\,($-1+i$) & \quad 0.8896052164\,($1-i$) \\
12 & \quad 0.9945374276E$-$12 & \quad 0.8896052164\,($1-i$) \\
\br
\end{tabular}
\end{indented}
\end{table}

\section{Resonances}

Poles of the scattering function are associated with what are called Gamow states: solutions of the Schr\"odinger equation, corresponding to complex energies $E=E_{\rm{R}}-i\Gamma/2$ ($\Gamma>0$), which at large distances contain only outgoing waves, as shown in Eq. (\ref{i41}). They owe their name to the fact that they were first used by Gamow \cite{gamo} to account for experimental data of $\alpha$-decay of certain nuclei. Due to the non-vanishing imaginary part of its energy, Gamow states are suitable to represent a decaying state, whose time evolution would be given by
\begin{equation}
\Psi(x,t)=\exp(-\,\Gamma\,t/2)\,\exp(-\,i\,E_{\rm{R}}\,t)\,\psi(x).   \label{i42}
\end{equation}

The question arises whether each one of these Gamow states may be associated to a resonance, that is, to an enhancement of the interaction with the potential at real energies in the neighborhood of $E_{\rm{R}}$. Resonances in a potential are characterized by a sudden increase of about $\pi$ in the phase shift as the energy increases. This occurs when a pole of $S(E)$ lies in the vicinity of the real $E$ axis. In this case, constant phase-lines converging at the pole and corresponding to values of $\arg S(E)$ in an interval of amplitude $\pi$ have their intersections with the real $E$ axis contained in a small interval of values of $E$. Looking at the modulus and phase chart shown in Fig. 3, we realize that the pole at $(0.889605, -0.889605)$ could be associated to a resonance. This is more clearly seen in Fig. 4, where we show the time delay suffered by a wave interacting with the potential Eq. (\ref{i4}) at energies in the interval $(-10, 15)$. In our case, the time delay, defined as \cite[pp 110--111]{nuss}
\begin{equation}
\Delta t=2 \hbar\,\frac{d\delta (E)}{dE},  \label{i44}
\end{equation}
turns out to be
\begin{equation}
\Delta t=2 \hbar\,\Im\left(\frac{dN(E)/dE}{N(E)}\right),  \label{i45}
\end{equation}
with $N(E)$ given in Eq. (\ref{i36}) and
\begin{equation}
\frac{dN(E)}{dE} = e^{i\pi/8}\, \frac{\psi\left(\frac{3-E}{4}\right)-i\,\psi\left(\frac{1+iE}{4}\right)}{4\,\Gamma\left(\frac{3-E}{4}\right)\,\Gamma\left(\frac{1+iE}{4}\right)}
+ e^{-i\pi/8}\,\frac{\psi\left(\frac{1-E}{4}\right)-i\,\psi\left(\frac{3+iE}{4}\right)}{4\,\Gamma\left(\frac{1-E}{4}\right) \,\Gamma\left(\frac{3+iE}{4}\right)},  \label{i46}
\end{equation}
where $\psi(\ldots)$ represents the digamma function.
 The marked peak in Fig. 4 reveals the mentioned resonance. There is also a much less marked bump at $E\approx 4$, obviously associated to the pole at $(4.081280, -2.977506)$. Other poles seem to have no physical implication.

\begin{figure}
\begin{center}
\vspace{0.5cm}\includegraphics{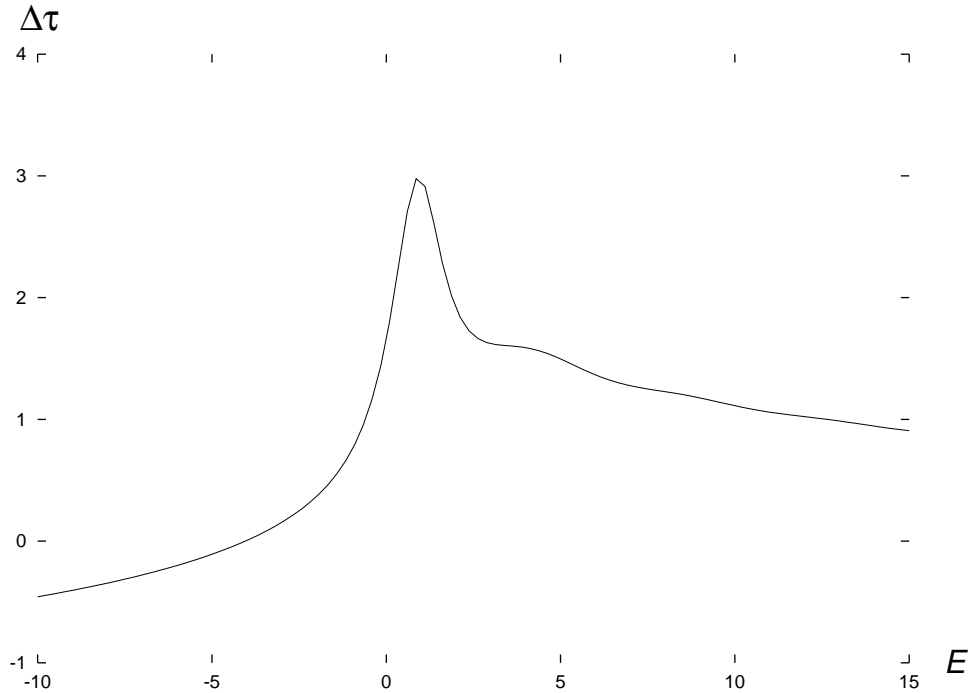}
\end{center}
\caption{Time delay of the outgoing wave as a function of the energy. The scale is consistent with that used for lengths and energies.}
\label{fig:4}
\end{figure}

For illustration, we present in Figs. 5 to 7 the square of the modulus of the wave function for three different values of the energy, namely,
$E_G=0.889605-0.889605\,i$ (Gamow state), $E_r=0.935$ (resonance, large time delay), and $E_s=-4.042626$ (saddle point, time delay equal to zero). In the first case the coefficients $A_1$ and $A_2$ in Eq. (\ref{i18}) are determined by Eq. (\ref{i27}) together with an arbitrarily chosen normalization condition
\[
A_1\,T_{1,3}^++A_2\,T_{2,3}^+=1,
\]
in such a way that
\[
\psi_{\rm{phys}}(x) \sim \psi_3^+(x), \qquad \mbox{for} \quad x\to +\infty.
\]
The (non-normalized) probability density shown in Fig. 5 corresponds to a time $t=0$. Subsequently it retains the same shape, but decreases, according to Eq. (\ref{i42}), by a factor $\exp (-\Gamma t)$, with $\Gamma=1.77921$. In the cases of Figs. 6 and 7, the coefficients $A_1$ and $A_2$ are obtained from Eqs. (\ref{i27}) and (\ref{i28}). As the energy is real, time damping does not occur. The oscillations in the value of the (non-normalized) probability density are due to the interference of the incoming and outgoing waves. The amplitude of the oscillations goes as $x^{-1}$ for $x\to +\infty$. In the case of resonance (Fig. 6), the large probability density at $x=0$ is to be noticed.
\begin{figure}
\begin{center}
\vspace{0.5cm}\includegraphics{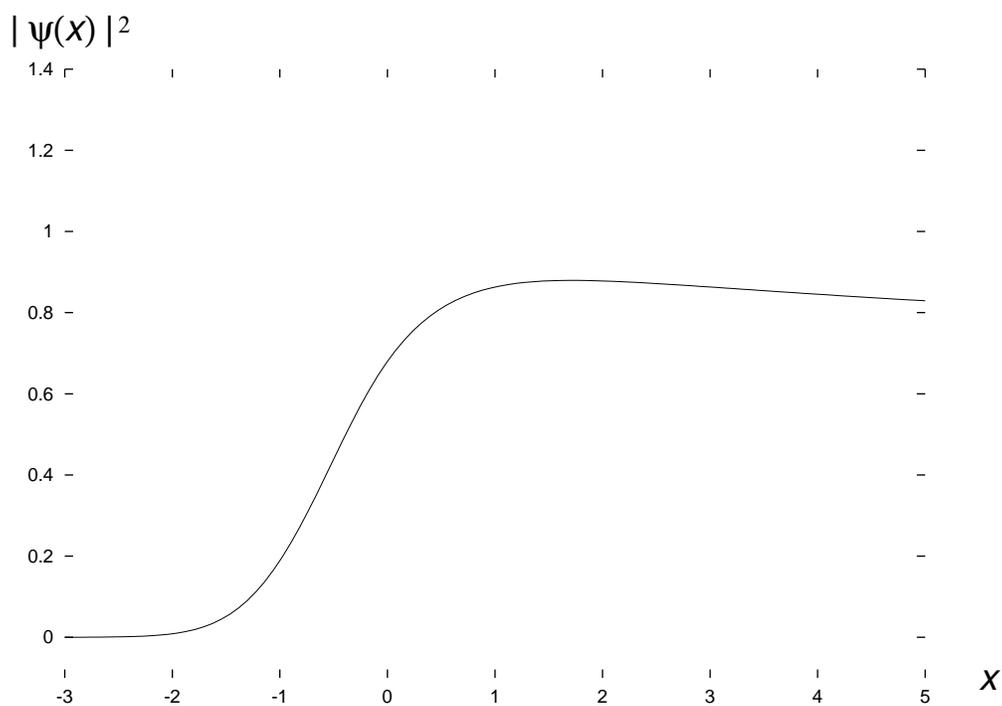}
\end{center}
\caption{Squared modulus of the wave function of the Gamow state of energy $E_G=0.889605-0.889605\,i$.}
\label{fig:5}
\end{figure}

\begin{figure}
\begin{center}
\includegraphics{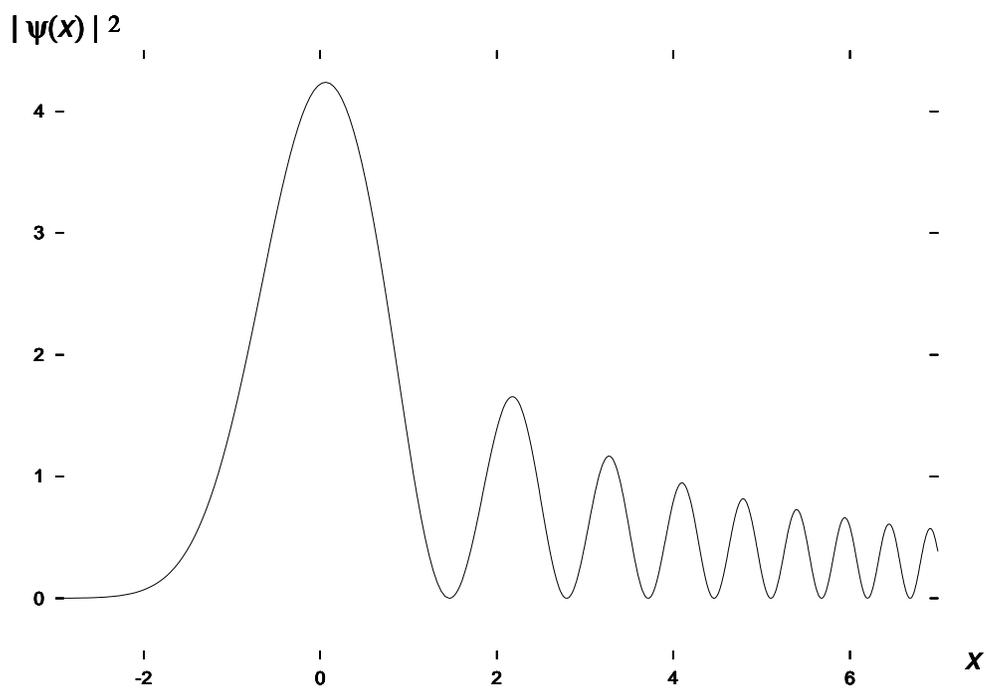}
\end{center}
\caption{Squared modulus of the wave function resulting by interference of the incoming and outgoing waves at the energy of resonance, $E_r=0.935$.}
\label{fig:6}
\end{figure}

\begin{figure}
\begin{center}
\vspace{0.5cm}\includegraphics{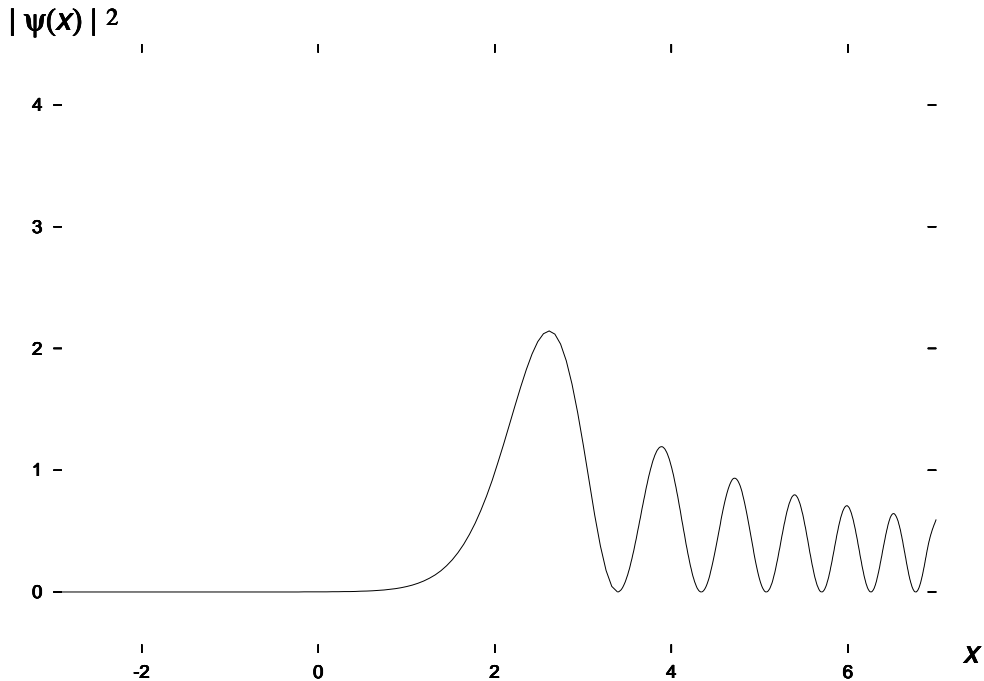}
\end{center}
\caption{Squared modulus of the wave function resulting by interference of the incoming and outgoing waves at energy $E_s=-4.042626$.}
\label{fig:7}
\end{figure}

\section{Final comments}

The parabolic odd potential, Eq. (\ref{i4}), considered in this paper is an unorthodox one. Usual treatment of scattering refers to three-dimensional spherically symmetric potentials of finite range. Concepts like cross-section, phase shifts, $S$ matrix, resonances, etc., are well established for those potentials. The idea of cross-section does not seem to be extendable to our one-dimensional potential. The other
concepts can be defined in a consistent and natural way, as we have seen. Nevertheless, the peculiar characteristics of the potential, namely being totally-reflecting, of infinite range, and unbounded, originate obvious differences with the usual three-dimensional case. Some comments about these differences are in order, we believe.

We have needed only one Riemann sheet to describe $S$ as a function of $E$. In the case of potentials of finite range, the elements of the $S$ matrix depend on $E$ through its square root, a bi-valued function. Two Riemann sheets, the so called physical and unphysical ones, are needed. For this reason it is preferable to express $S$ in terms of the wave number $k$, proportional to the square root of $E$. In our case, a global definition of wave number is neither possible nor necessary.

Gamow states in a finite range spherically symmetric potential have a complex wave number $k$ whose imaginary part is negative. This implies that the outgoing wave, $\exp(ikr)$, increases exponentially with $r$, a property which could be considered unreasonable. However, in words of Garc\'{\i}a-Calder\'on and Peierls, \cite{garc} ``such increase is entirely reasonable, because it reflects the fact that we are assuming an exponentially decaying state, and thus we see at distance $r$ the particles emitted by the system a time $r/v$ earlier, where $v$ is their velocity, and these are more numerous by a factor $\exp (r/v\tau)$; $\tau$ being the mean life." In the parabolic odd potential, the probability density, as shown in Fig. 4, does not increase with $x$, but, according to Eq. (\ref{i11}), it goes as $x^{-0.110395}$ for $x\to +\infty$. Such a behavior is consistent with the fact that the (local) wave number, or, in other words, the velocity of the outgoing particle represented by the wave function increases with $x$. Thinking in terms of a classical particle, the time needed to reach a large distance $x\to +\infty$ goes as $x$ in the case of a potential of finite range, whereas it goes as $\log x$ in our case: the particle escapes much more rapidly in the parabolic potential.

Obviously, the potential considered in this paper is an idealization. Its interest lies mainly in the fact that closed analytical forms can be obtained for the solutions of the Schr\"odinger equation (Eqs. (\ref{i9}) to (\ref{i12}) and (\ref{i14}) to (\ref{i17})), the scattering function (Eq. (\ref{i35})), and the time delay (Eq. (\ref{i45})). Nevertheless, given the continuous progress in the synthesis of artificial quantized structures by means of stacks of thin films, the possibility of the odd parabolic potential to represent a useful approximation to a real situation should not be discarded.

\ack{The comments of three anonymous referees have considerably contributed to improve the presentation of this article.
Financial support from Conselho Nacional de Desenvolvimento
Cient\'{\i}fico e Tecnol\'{o}gico  (CNPq, Brazil) and from Departamento de Ciencia, Tecnolog\'{\i}a y Universidad del Gobierno de Arag\'on (Project E24/1) and Ministerio de Ciencia e Innovaci\'on (Project MTM2009-11154) is gratefully acknowledged.}


\section*{References}

\begin{thebibliography}{99}

\bibitem{yari} Yaris R, J. Bendler J, Lovett R A , Bender C M and Fedders P A 1978 {\it Phys. Rev. A} {\bf 18} 1816
    \nonum Caliceti E, Graffi S and Maioli M 1980 {\it Commun. Math. Phys.} {\bf 75} 51

\bibitem{alva1} Alvarez G 1988 {\it Phys. Rev. A} {\bf 37} 4079

\bibitem{alva2} Alvarez G 1989 {\it J. Phys. A: Math. Gen.} {\bf 22} 617
     \nonum Alvarez G 1995 {\it J. Phys. A: Math. Gen.} {\bf 27} 4589
     \nonum \'Alvarez G and Casares C 2000 {\it J. Phys. A: Math. Gen.} {\bf 33} 2499
     \nonum \'Alvarez G and Casares C 2000 {\it J. Phys. A: Math. Gen.} {\bf 33} 5171

\bibitem{jent1} Jentschura U D, Surzhykov A, Lubasch M and Zinn-Justin J 2008 {\it J. Phys. A: Math. Theor.} {\bf 41} 095302
    \nonum Jentschura U D and Zinn-Justin J  2010 {\it Appl. Numer. Math.} {\bf 60} 1332

\bibitem{bart} Barton G 1986 {\it Ann. Phys.} {\bf 166} 322

\bibitem{bala} Balazs N L and Voros A 1990 {\it Ann. Phys.} {\bf 199} 123
    \nonum Castagnino M, Diener R, Lara L and G. Puccini G 1997 {\it Int. J. Theor. Phys.} {\bf 36} 2349
    \nonum Shimbori T and Kobayashi T 2000 {\it Nuovo Cimento} {\bf 115 B} 325

\bibitem{land} Landau L D and Lifshitz E M 1958 {\it Quantum Mechanics} (Reading, MA: Addison-Wesley)

\bibitem{robi} Robinett R W 1997 {\it Quantum Mechanics} (Oxford: Oxford University Press)

\bibitem{newt} Newton R G 2002 {\it Quantum Physics} (New York: Springer)

\bibitem{boya} Boya L J 2008 {\it Riv. Nuovo Cimento} {\bf 31} 75

\bibitem{ding} Dingle R B 1973 {\it Asymptotic Expansions: Their Derivation and Interpretation} (London: Academic Press)

\bibitem{kahn} Kahn A H 1961 {\it Am. J. Phys.} {\bf 29} 77
    \nonum Eberly J H 1965 {\it Am. J. Phys.} {\bf 33} 771
    \nonum Form\'anek J 1976 {\it Am. J. Phys.} {\bf 44} 778

\bibitem{wren} Wrench J W (1968) {\it Math. Comp.} {\bf 21} 617

\bibitem{abad} Abad J and Sesma J (2003) {\it Comput. Phys. Comm.} {\bf 156} 13

\bibitem{gamo} Gamow G 1928 {\it Z. Phys.} {\bf 51} 204

\bibitem{nuss} Nussenzveig H M 1972 {\it Causality and Dispersion Relations} (New York: Academic Press)

\bibitem{garc} Garc\'{\i}a-Calder\'on G and Peierls R 1976 {\it Nucl. Phys.} {\bf A265} 443

\end{thebibliography}
\end{document}